# Evolution of Intrinsic Vacancies and Prolonged Lifetime of Vacancy Clusters in Black Phosphorene

Yongqing Cai[a,b,#], Shuai Chen[b,#], Junfeng Gao[c], Gang Zhang[b,*], Yong-Wei Zhang[b,*]

[a]*Joint Key Laboratory of the Ministry of Education, Institute of Applied Physics and Materials Engineering, University of Macau, Taipa, Macau, China*

[b]*Institute of High Performance Computing, A\*STAR, Singapore 138732*

[c]*Key Laboratory of Materials Modification by Laser, Ion and Electron Beams, Dalian University of Technology, Dalian, China 116024*

*Corresponding author: zhangg@ihpc.a-star.edu.sg; zhangyw@ihpc.a-star.edu.sg*

*#These authors contribute equally*

## Abstract

Due to the relatively low formation energies and highly mobile characteristics of atomic vacancies in phosphorene, understanding their evolutions becomes crucial for its structural integrity, chemical activities and applications. Herein, by combining first-principles calculations and kinetic Monte Carlo simulation, we investigate the time evolution and formation of atomic vacancy clusters from isolated monovacancies (MVs), aiming to uncover the mechanisms of diffusion, annihilation, reaction of these atomic vacancies. We find that while isolated MVs possess a highly mobile character, they react and form MV pairs which possess much lower mobility and high stability under ambient condition. We also show that the disappearance of MVs at the edge is quite slow due to the relatively high energy barrier, and as a result, around 80% of MVs remains even after two years under ambient condition. Our findings on one hand provide useful information for the structural repairing of phosphorene through chemical functionalization of these vacancy clusters, and on the other hand, suggest that these rather stable vacancy clusters may be used as an activated catalyst.

## Introduction

Phosphorene, a two-dimensional monolayer of layered black phosphorus, has attracted significant interest in optoelectronics and photonic applications.[1-4] It also holds great promise for solar energy utilization[5-10] owing to its robust direct band-gap for all thicknesses.[11] Its puckered structure and strong interlayer coupling induce highly anisotropic electronic and phononic properties, allowing facile ways to manipulate the light-matter interaction and thermal transport.[12,13] Recently, through forming hetero-interfaces with other functional semiconductors, such as graphitic carbon nitride or CdS, phosphorene was found to have a good activity as a metal-free photocatalyst for hydrogen evolution reaction from visible to near-infrared range.[14-17]

The relatively weak P-P bonding and rich hybridized states of P atom enable phosphorene to have many unique properties, such as high structural flexibility due to its low in-plane and out-of-plane stiffness, high atomic vacancy concentrations due to low formation energy, diverse polymorphs due to multiple coordination numbers. Our previous study revealed a nearly itinerant behavior of vacancies within atomically thin phosphorene plane even at room temperature (RT), which is in contrast to its renowned predecessor, graphene, which possesses an ultralow vacancy mobility, but an ultrahigh electron/hole mobility.[18] In general, the presence of defects can alter the electronic property[19] and chemical activity of a material.[20] Interestingly, while phosphorene shows a poor stability upon exposure to air, which was generally ascribed to the effect of water and oxygen, recent experimental studies showed that phosphorene could be exfoliated and remain stable in oxygen-free condition like liquid solvents[21,22] deoxygenated water.[23,24]

In contrast to graphene in which the concentration of thermally activated vacancies is low and thermally activated motion of atomic vacancies is rare at the temperature range of 300~ 600 K, [25-27], phosphorene possesses both low vacancy formation energy and low vacancy diffusion barrier and thus serves as an ideal material for examining the evolution of intrinsic vacancies or glass behavior in the 2D limit. For example, continuous annealing of phosphorene in vacuum at 400 °C leads to an amorphous red P like skeleton[28], which is suspected to be an effect of complex interactions of atomic vacancies in phosphorene. Also, recent experiment demonstrated a top-down route for the manipulation of isolated atoms or vacancies in phosphorene structures for production of atomic scale structures via electron beam sculpturing inside a transmission electron microscope.[29] These experimental studies revealed rich nanoscale defect kinetics triggered by moderate electric or thermal field. However, due to the nearly itinerant nature of atomic vacancies, the intrinsic features related to the nucleation, migration and coalesce of vacancies are largely unknown. This is because, it is still a significant challenge to experimentally capture the processes of the reconstruction of isolated monovacancies (MVs) and their reactions. In particular, it remains unclear under what conditions two isolated MVs are able to react to a divacancy (DV) or a MV pair, and what conditions three isolated MVs are able to form a trivacancy (TV) or a vacancy cluster. Such information, which is difficult obtained through traditional experiments due to the difficulty in quantitatively pinpointing and recording the varying positions of the atomic vacancies (either MV or vacancy clusters) in experiments, is critical for the stability and performance of phosphorene-based devices.

In this work, by using first-principles calculations and kinetic Monte Carlo simulations, we investigate the atomic scale kinetics of atomic vacancies with respect to the annihilation and evolution. The time-dependent evolutions of small vacancy clusters, including MV, DV, and TV, are simulated, showing the formation of bigger-size vacancy clusters and explaining possible underlying mechanism for the formation of voids observed in experiment. Moreover, we examine energetics and kinetics of atomic vacancies in the proximity of the edge, aiming to understand the annihilation likelihood of vacancies near the edge. Our work provides insights into the time-dependent evolution of the atomic vacancies, which may be useful for preparing and characterizing phosphorene nanostructures and understanding the stability of phosphorene based devices.

## COMPUTATIONAL METHODS

**First-principles Quantum Simulations:** We perform first-principles calculations by using Vienna ab initio simulation package (VASP) within the framework of density functional theory (DFT) to obtain the energetics and kinetics of the atomic vacancies.[30] The Perdew–Burke–Ernzerhof (PBE) functional is selected as the exchange-correlation functional. Atomic vacancies are modelled by creating a 5×4 supercell of monolayer phosphorene based on the relaxed unit cell with lattice constants of 3.305 and 4.617 Å. A vacuum layer of thickness greater than 15 Å is adopted. A 3×3×1 Monkhorst-Pack grid and a kinetic energy cutoff of 400 eV are used. The atomic structures are fully optimized until the atomic forces are less than 0.005 eV/Å.

**Kinetic Monte Carlo simulations on Evolution of Vacancy:** Kinetic Monte Carlo (kMC) simulations are performed with atomic models where the randomly-distributed vacancies are constructed. In the models, the total number of lattice sites is 40000, corresponding to a lateral size of 41.1×32.8 nm$^2$. The boundary conditions are periodic, and the vacancies only occupy or diffuse across the lattice sites (on-lattice simulation). At each given state $i$, kMC model specifies all the possible states $j$ that they may transit into at the next step. The transition rate is calculated by the transition state theory[31]: $r_{i \to j} = \nu \exp(-\frac{E_{i \to j}}{k_b T})$, where $\nu$ is the frequency of atomic vibration (~ $10^{12}$ s$^{-1}$), $E_{i \to j}$ is the energy barrier for the transition from current state $i$ to another state $j$, $k_b$ is the Boltzmann constant, and $T$ is the temperature (300 K). The events considered in the model are the MV diffusion along zigzag ($E_{MV-D-ZZ}$=0.3 eV) and armchair edges ($E_{MV-D-AC}$=0.5 eV), MV pair/cluster formation along zigzag ($E_{MVx-F-ZZ}$=0.3 eV) and armchair edges ($E_{MVx-F-AC}$=0.5 eV), MV pair/cluster diffusion/break along zigzag ($E_{MVx-D-ZZ}$=1.13 eV) and armchair edges ($E_{MVx-D-AC}$=1.13 eV), DV/TV+ (formation from equal or more than two MVs) formation along zigzag ($E_{DV-F-ZZ}$=1.44 eV) and armchair edges ($E_{DV-F-AC}$=1.44 eV), and DV/TV+ break along zigzag ($E_{DV-B-ZZ}$=2.44 eV) and armchair edges ($E_{DV-B-AC}$=2.44 eV). kMC simulations with free boundary conditions are performed for phosphorene nanoribbons, where the escaping barrier for MV at the edge, 0.5 eV, is used. Note that all the energy barriers are obtained from first-principles calculations.[18] It should be noted that the coalescence of MV$_x$ clusters involves the sliding and shifting of phosphorus atoms along zigzag chain, which have a maximum barrier of 0.3 eV (see Ref. 18), therefore the energy barrier of 0.3 eV is used for the KMC simulation of MV$_x$ clusters.

**Results and Discussions:**

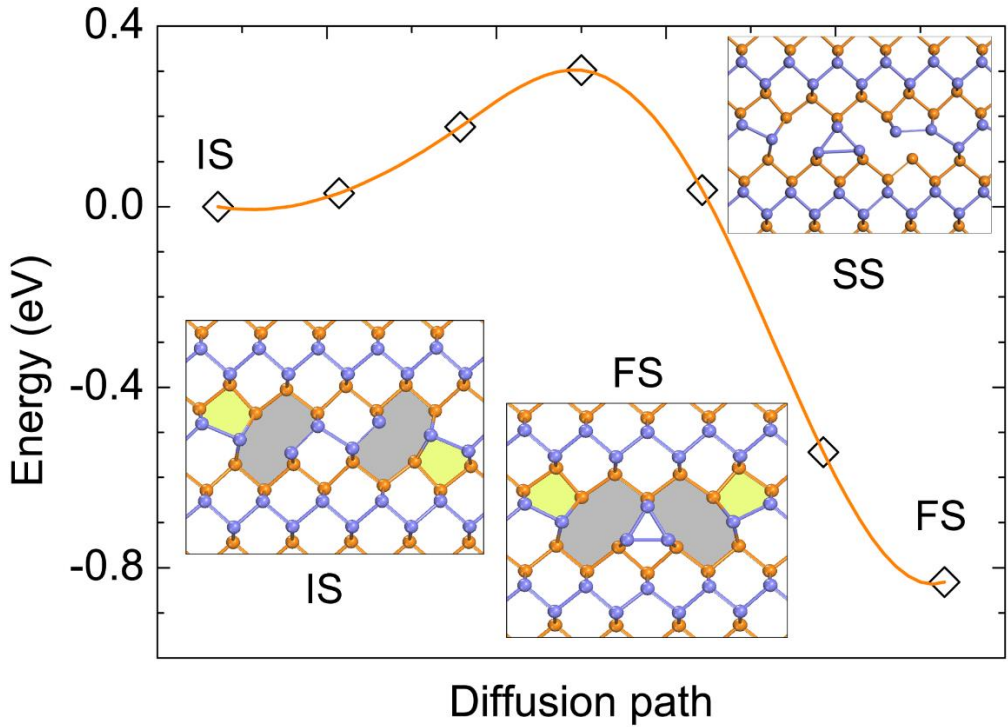

**Figure 1.** Coalescence of two MVs along a zigzag chain. Activation energy of the formation of a MV pair as the final state (FS), which is formed by coalescing two well separated MVs as the initial state (IS) through the saddle state (SS).

Energetic and kinetic behaviors of single MV and DV were well studied before.[18] It was shown that these vacancies exhibit a highly mobile behavior even through sole thermal excitation.[18] Therefore, at a high concentration, the probability of interaction of isolated MVs increases significantly. In this work, we firstly examine the kinetics of the pairing process of two isolated MVs. Since a MV hops much more frequently (around thousands faster) along the zigzag direction than along the armchair direction,[18] herein we consider the colliding situation along the zigzag direction. As shown in Figure 1, our DFT claculations show that two isolated MVs prefer to combine to form a MV pair ($MV_2$), which is an exothermic process with a small barrier of 0.3 eV (Figure 1), rather than a DV, with the latter process exhibiting a much higher energy barrier of 1.4 eV (see Ref. 18). The reverse process, splitting of the MV$_2$ pair into two isolated MVs, requires a barrier of around 1.1 eV. Therefore, the mobility of a MV$_2$ is rather limited. For a phosphorene sheet with a high concentration of MVs, while the inner MVs (far from the edges) are highly mobile, their mobility drops significantly once forming MV pairs, which are largely retained or pinned due to a high diffusion energy barriers of the latter.

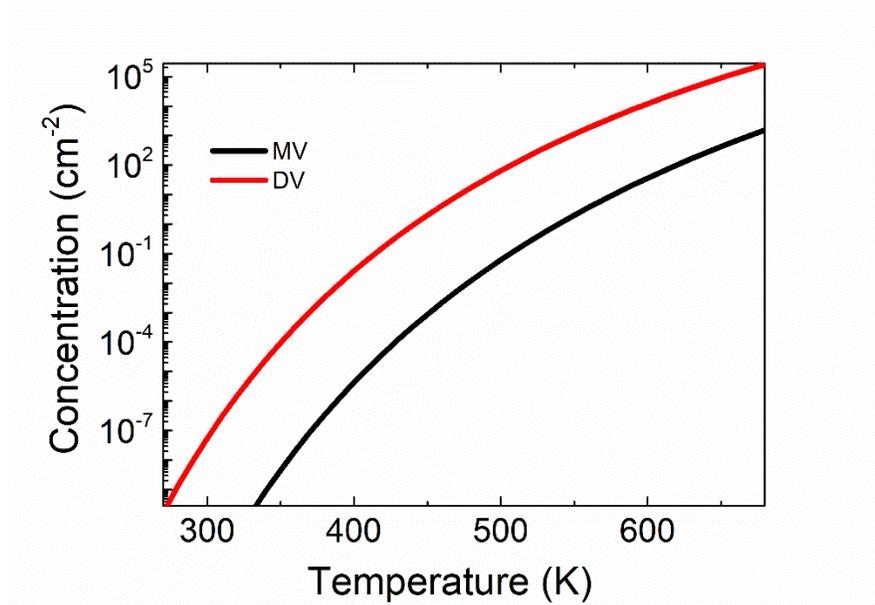

**Figure 2**. Estimated concentrations of MV and DV in phosphorene as a function of temperature.

At finite temperature $T$, the average equilibrium concentrations of MV and DV in monolayer phosphorene can be estimated by $n/N = \exp(-\Delta E / k_B T)$, where $n$ is the number of the defect, $N$ is the total number of atomic sites in phosphorene, $k_B$ is the Boltzmann constant, and $\Delta E$ is the formation energy with a value of 1.65 eV for MV and that of 1.35 eV for DV[18]. Although the thermodynamic equilibrium concentrations of both MV (~$10^{-12}$ cm$^{-2}$) and DV (~$10^{-7}$ cm$^{-2}$) are low at room temperature, their concentrations increase rapidly with increasing the temperature (Figure 2). Alternatively, high concentrations of defects can also be achieved by irradiation, which is a well-accepted approach for the introduction and control of defects in graphene.[32,33] It is noted that the formation energies of MV (1.65 eV) and DV (1.35 eV) in phosphorene are much lower than those in graphene (7.57 eV for MV and 8.09 eV for DV).[18] Therefore, introducing vacancies into phosphorene should be much easier than that into graphene. A previous theoretical study[34] predicted that MVs or vacancy clusters can be formed in phosphorene under 80 keV electron beam. Several experiments have been demonstrated by using atomic scale engineering on the structures of phosphorene via electron irradiation. For instance, preparation of phosphorus atomic chains in black phosphorus via electron beam sculpturing was demonstrated by Xiao et al.,[29] and chain vacancy in black phosphorous was created by electron-beam irradiation.[35] Clearly, controlled structural modification of few-layer phosphorene with sub-nanometer precision can be achieved by sub-nanometer-precise scanning transmission electron microscopy (STEM) nanosculpting.[35]

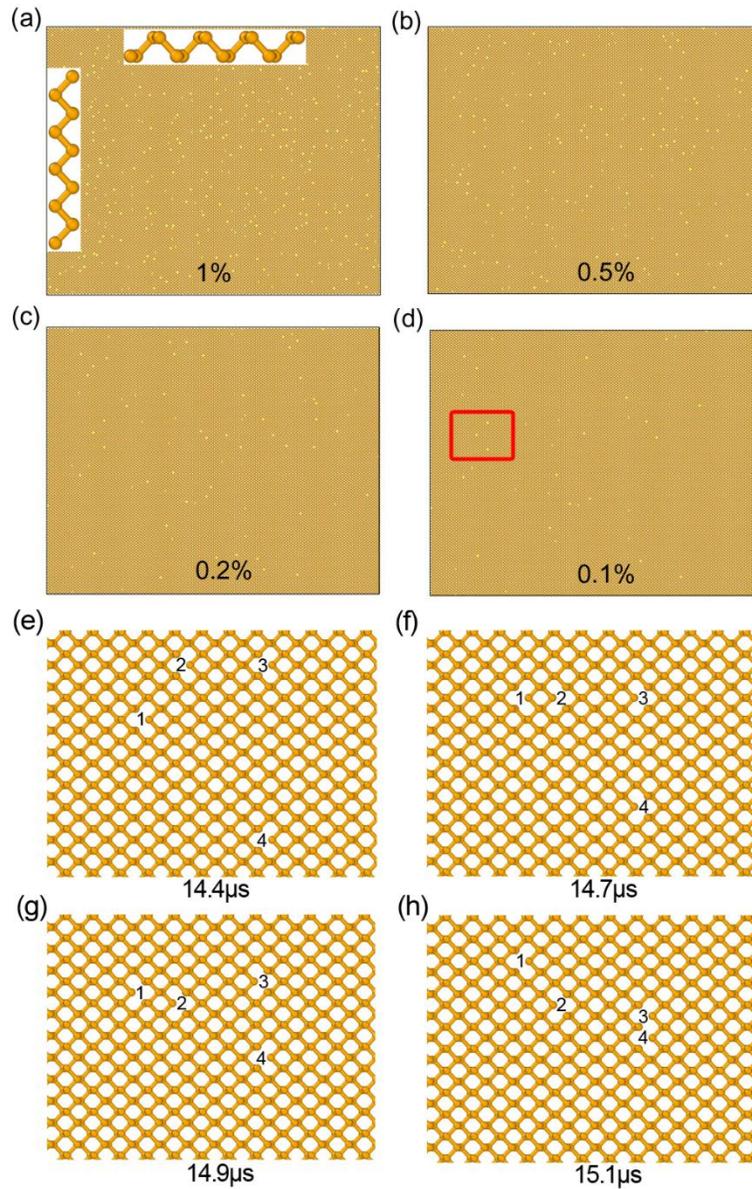

**Figure 3**. Initial lattice models with four different vacancy concentrations: (a) 1%, (b) 0.5%, (c) 0.2%, and (d) 0.1%. (e-h) Snapshots of the region marked in (d) during the evolution process of vacancies.

An important feature of MV of phosphorene is with regard to its highly mobile character. To gain quantitative understanding of the time evolution behavior and kinetic processes of vacancies, we performed kMC simulations with energetics obtained from first-principles calculations. Here, we considered four initial models with different concentrations of randomly distributed MVs: 1%, 0.5%, 0.2% and 0.1% (Figure 3a to d). The zigzag and armchair directions are marked in Figure 3a. Snapshots of the region labeled in Figure 3d during the evolution process of vacancies are shown in Figure 3e to h. It indicates that the MVs (tagged as 1 to 4) always diffuse (Figure 3e to g) and coalesce (Figure 3h) along zigzag directions. The diffusion and coalescence processes of vacancies in phosphorene show a strong anisotropic behavior, which is very different from those in graphene[32] and silicene.[36] The MVs favor aggregations by pairing or forming MV clusters (MV$_x$ with $x$ being the number of MV) mostly along zigzag direction. Similar to the motion of single MV, the diffusion, coalescence and splitting of MV clusters mainly involve the neighboring-site and on-site hopping, which are made possible by the diffusion of phosphorus atoms along the zigzag chains. The energy barriers for the diffusion process of MV and aggregation process of MV$_2$ are only about 0.3 eV along zigzag direction (Figure 1), which is lower than the energy barrier for MV escaping from the edge (0.5 eV). The occurrence rate for MV diffusion/aggregation is $9.1 \times 10^6$ s$^{-1}$ at 300K, which is three orders higher than that of MV escaping from the edge at 300K ($4.0 \times 10^3$ s$^{-1}$). After aggregation, the motion of MV$_x$ with an energy barrier of 1.3 eV is, however, almost pinned at room temperature. Because of the ultralow barrier of 0.3 eV for coalescence and merging, there is no limitation of $x$ for the aggregation process of MV$_x$ in the above analysis.

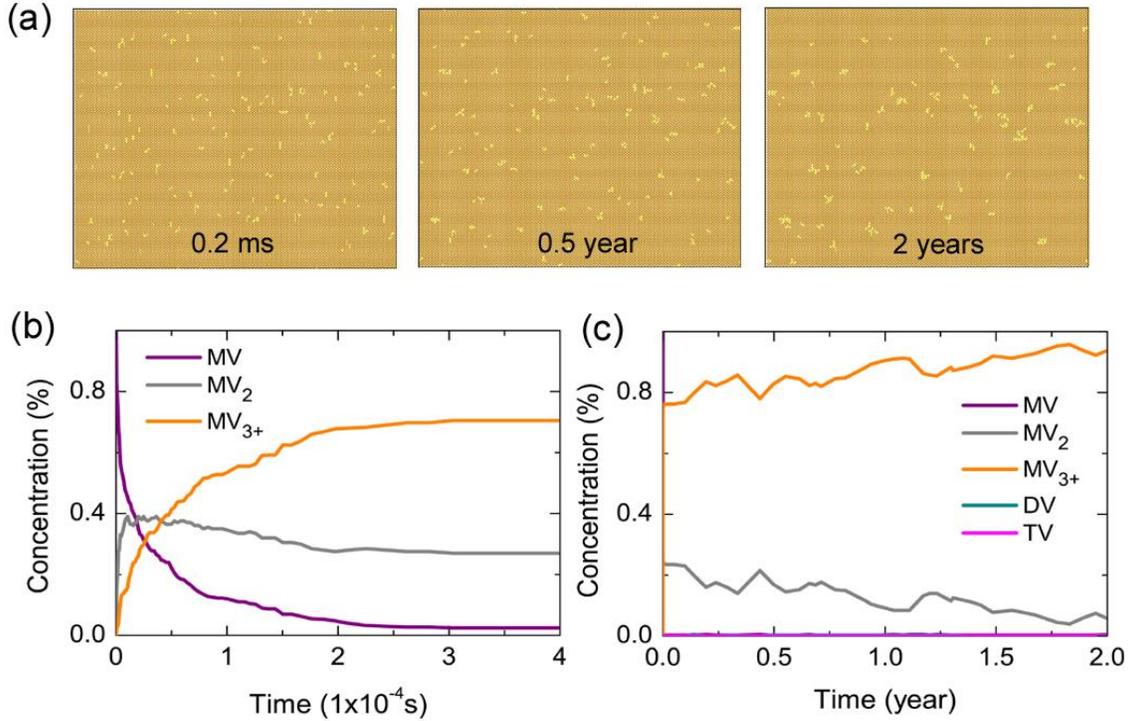

**Figure 4**. Evolution of vacancies with time in phosphorene. (a) Snapshots of the ensemble in the evolution process of vacancies by kMC method. Variations of the concentrations of vacancies within (b) short time ($10^{-4}$ s) and (c) long time (year). The total vacancy concentration is 1%. $MV_2$: MV pair; $MV_{3+}$: cluster containing equal to or more than three MVs; DV: divacancy; and TV: trivacancy.

Figure 4a shows the snapshots of the ensemble in the evolution of vacancy at the initial concentration of 1% MVs. The variations of vacancy concentrations at the short-time and long-time evolutions are shown in Figure 4b and c, respectively. It is seen from Figure 4b that the MVs diffuse fast and form MV pairs and clusters within ~0.01 ms initially, leading to the dramatic decrease of MV concentration and increase of MV pair and cluster concentrations. After the MV pair concentration reaches a peak value (~0.4%) at 0.01 ms, the MVs have a high probability to coalesce with MV pairs to form MV clusters, resulting in the decrease of MV pair concentration and continuous increase of MV cluster concentration from 0.01 ms to 0.2 ms. Subsequently after 0.2 ms, the evolution slows down significantly because only a small amount of MVs are left over. Within the next 2 years, the MV pairs break into two MVs occasionally (Figure 4c). The newly formed MVs further coalesce with MV pairs to become MV clusters, resulting in the slow decrease of MV pair concentration and increase of MV cluster concentration. However, the MV pairs and clusters have a quite high energy barrier to evolve into DV and TV (1.44 eV). Therefore, high concentrations of MV pairs and clusters are mostly retained and only very small amounts of DVs and TVs are formed (Figure 4c).

Next we present our theoretical analysis of evolution of vacancies based on rate equations. Analytical rate equations based on the energy barriers along zigzag edges are carried out since the diffusion of MV along zigzag edge is significantly faster than that along armchair edge. The rate equations for the concentrations of MV ($C_1$), MV pair ($C_2$) and MV cluster ($C_3$) are:

$$dC_1/dt = -2C_1^2 \times r_F - C_1 \times C_2\, r_F + 2C_2 \times r_B + C_3 \times r_B \qquad (1)$$

$$dC_2/dt = 2C_1^2 \times r_F - 2C_1 \times C_2\, r_F - 2C_2 \times r_B + 2C_3 \times r_B \qquad (2)$$

$$dC_3/dt = 3C_1 \times C_2\, r_F - 3C_3 \times r_B \qquad (3)$$

where $r_F$ and $r_B$ are the occurrence rates for the formation and the breaking of MV pair/cluster along zigzag edges, respectively, which are calculated by the transition state theory with the corresponding energy barriers ($E_F$=0.3 eV, $E_B$=1.13 eV). The calculation results on the time evolution of vacancy concentrations are shown in Figure 5. It is seen that the MV concentration decreases dramatically and MV pair and cluster concentrations increase accordingly in the initial 0.01 ms. Subsequently from 0.01 ms to 0.1 ms, MV and MV cluster concentrations continuously increase but MV pair concentration starts to decrease. After 0.1 ms, the evolution slows down significantly. These calculations results based on theoretical rate equations (Figure 5) are in good agreement with the kMC simulations (Figure 4b), thus supporting the extremely slow evolution of the vacancy concentration in the long run.

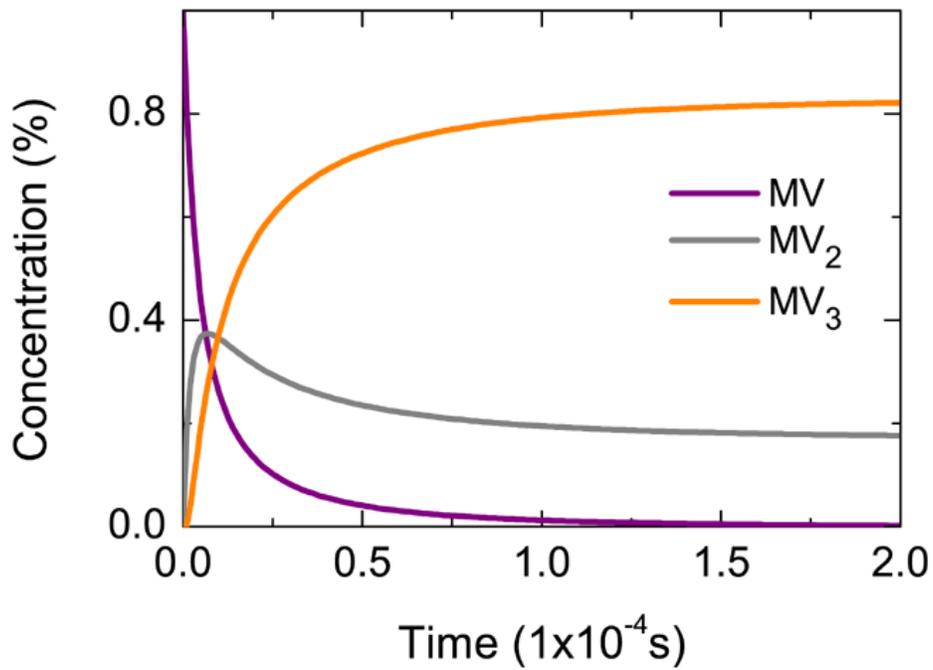

**Figure 5**. Rate equation analysis within the short time period. The initial total MV concentration is 1%.

To further verify our analysis and conclusions, more simulations with different initial vacancy concentrations (0.5%, 0.2% and 0.1%) are also carried out. Figure 6a, c and e show the snapshots of the ensemble at the time interval of 0.5 year when the initial total MV concentration is 0.5%, 0.2%, and 0.1%, respectively. Figure 6b, d and f plot the time evolution of vacancy concentrations, which exhibit a similar trend as that of 1% as shown in Figure 4. The MVs initially diffuse fast and form MV pairs and clusters in milliseconds. Subsequently, the MV pairs and clusters evolve very slowly in years. This slow-evolution state with a high concentration of MV pairs and clusters can sustain for years, signifying the high stability of MV pairs and clusters.

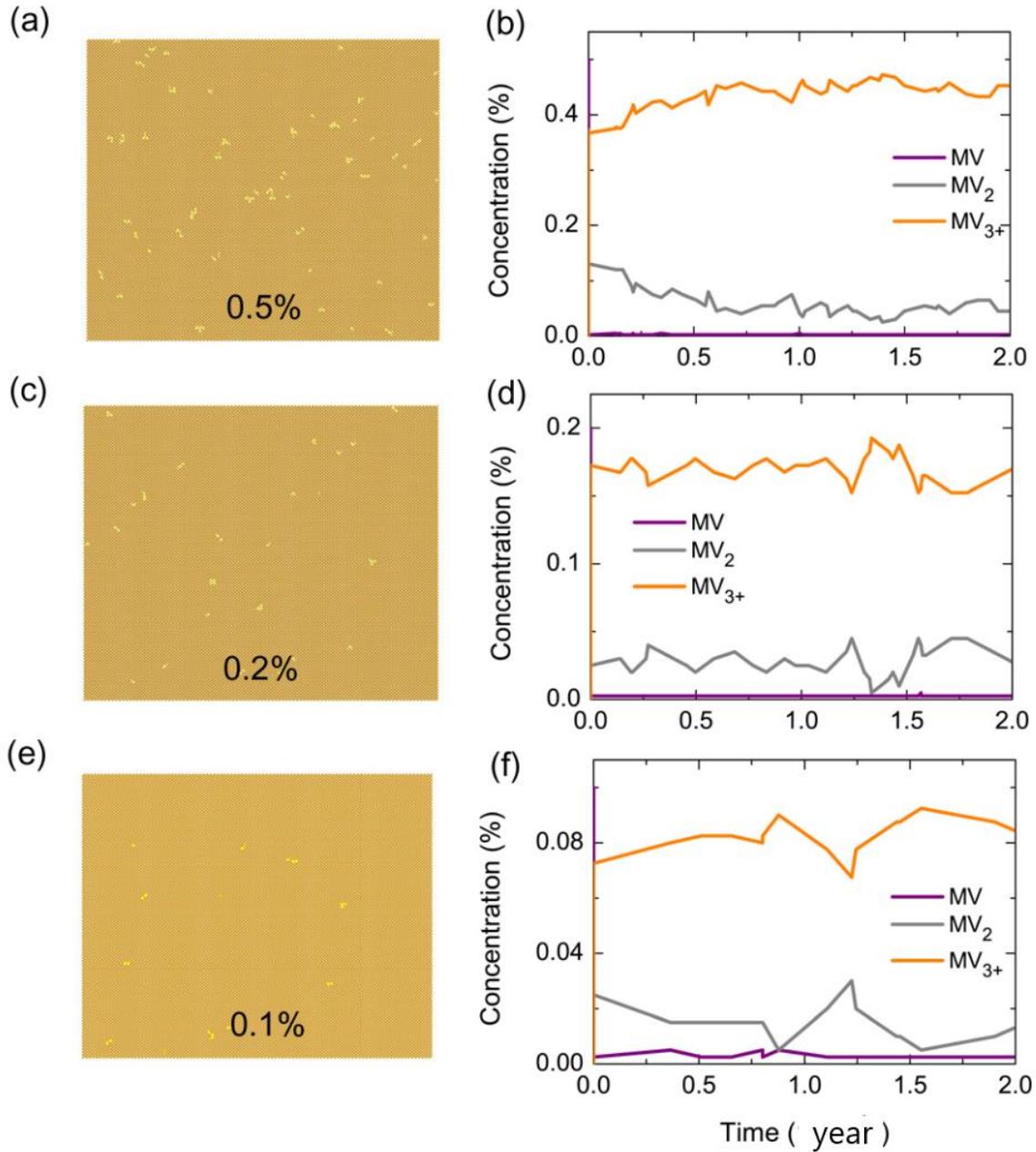

**Figure 6**. Snapshots of the ensemble at the time interval of 0.5 year when the total vacancy concentration is (a) 0.5%, (c) 0.2%, and (e) 0.1%. Variations of different kinds of vacancies with time when the total initial MV concentration is (b) 0.5%, (b) 0.2%, and (f) 0.1%. $MV_2$: MV pair; $MV_{3+}$: cluster containing more than three MVs.

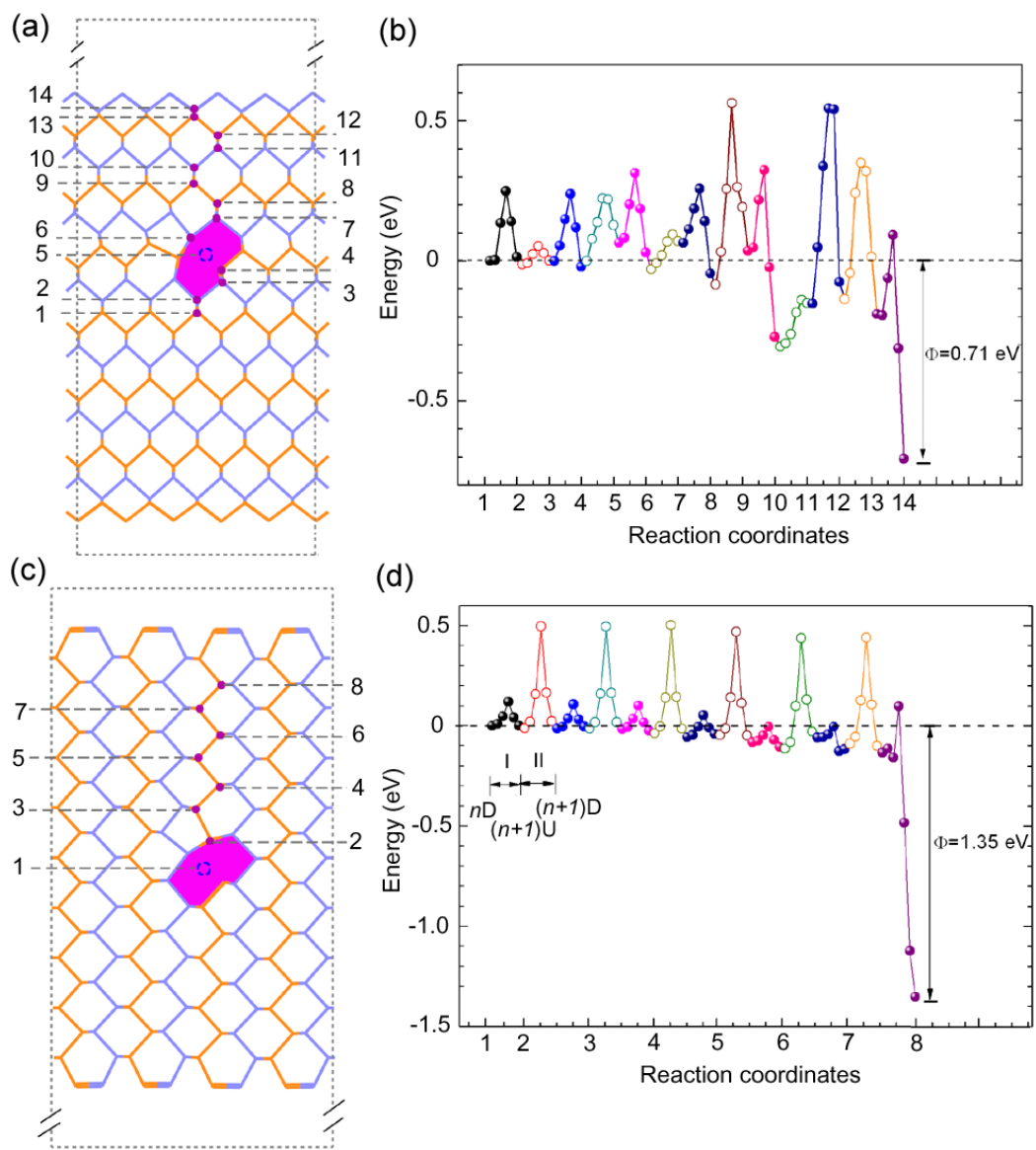

**Figure 7.** Interplay of the edges and MV of phosphorene. (a, c) Schematic of atomic models for the zigzag (a) and armchair (c) nanoribbons with the position of the MV marked by the numbers. (b, d) The activation barrier for the MV migrating from the interior to the zigzag (b) and armchair (d) edge by NEB calculation.

Finally, we investigate the interplay between the edges and MVs. Isolated MVs were found to diffuse with a small barrier in the sheet.[18] When the presence of edges is accompanied with rich defective states, these highly localized states may pin the Fermi level of the system and trap the MVs. By using the climbing nudged elastic band (NEB) method, the activating barriers are calculated for the MV moving from atomic sites in the interior to the edge along armchair (Figure 7a, b) and zigzag (Figure 7c, d) directions. Edge reconstruction was observed when vacancies were located exactly at the edge or the neighboring positions of the edge atoms. To avoid edge reconstruction, we only examined MV located near the edge from sites 1 to 8 marked in Figure 7c.

Starting from the central part of the ribbon (denoted as site 1), diffusion of a MV from site $n$ to a neighboring equivalent site $n+1$ along armchair direction has a maximum barrier of ~0.3 eV from site 1 to site 8. In the proximity of the edge, a larger barrier is found, ~ 0.7 eV from site 11 to site 12. For the diffusion along zigzag direction, similar to the diffusion in the 2D sheet,[18] there involve two basic steps: a neighboring-site hopping (I), and an on-site reconstruction (II). In both steps, the nonagon ring switches between up and down configurations, abbreviated as the "U" and "D", respectively, in Figure 7d. The maximum barrier is ~0.5 eV, which is found to be less dependent on the site of the intermediate MV. Therefore, our results indicate that MV has a relatively low energy barrier for migration near the edge, which is the same as the low barrier in 2D sheet. Compared to the initial configuration for MV at site 1 with its energy being set to zero, the various intermediate states are found to have a lower energy when the MV moves toward the edges, with a total energy drop ($\Phi$) of 0.71 and 1.35 eV for cases of zigzag edge and armchair edge, respectively.

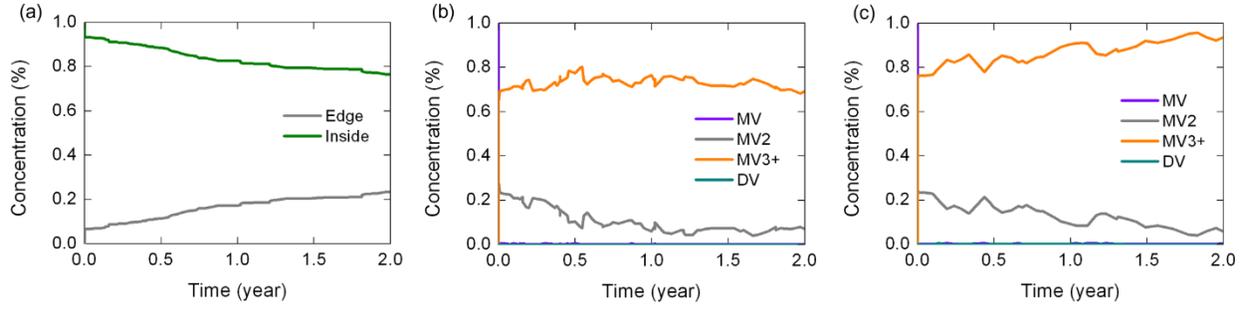

**Figure 8**. (a) Concentration of vacancies inside the domain and annihilation at the edge with free boundary condition. Concentrations of different types of vacancies with (b) free boundary condition and (c) periodic boundary condition.

Next we performed kMC simulation with free boundary condition for phosphorene nanoribbons, where the initial total vacancy concentration of MVs is 1% and the events of MV diffusion to and annihilation at the edge are considered. The variations of the vacancies within the sheet and annihilation at the edge with time is recorded and shown in Figure 8. It is seen that there is still 80% of vacancies remaining inside the domain after two years, and the variation trend of each type of vacancy with free boundary condition, which is shown in Figure 8b, is comparable to that with periodic boundary condition (Figure 8c).

Previous study predicted a rather mobile nature of MVs in phosphorene.[18] An interesting question is whether these MVs can diffuse to and annihilate at the edges or they can be retained within the sheet. Below we would like to show that the latter is likely to be the case. Our calculated energy barrier for MVs to escape at the edge (0.5 eV) is much higher than the diffusion barrier within the sheet (0.3 eV). This leads to the ratio of the escaping rate ($r$) at the edge to the diffusion rate within the sheet to be 1:2290 based on $r=ve^{-E/kT}$, where $v$ is the characteristic frequency, $E$ is the corresponding energy barrier and $T$ is the temperature (300 K). Our kMC calculations for phosphorene nanoribbons with free boundary conditions reveal that there is still 80% of vacancies within the sheet after two years (see Figure 8). Therefore, MVs only have a very limited chance to escape at the edge. The above result is also supported by low-temperature scanning tunneling microscopy/spectroscopy (STM/STS) experiments, which reported a large amount of MVs well-distributed in phosphorene sheet.[37] In addition, a high density of phosphorus vacancies distributed throughout the BP lattice was also observed experimentally.[38] It should be noted that the energy barrier for MVs to escape towards the edge (0.5 eV) is higher than the diffusion barrier within the sheet (0.3 eV). This difference would affect the evolution and distribution of vacancies at low and moderate temperatures. However, with increasing the temperature, the energy barriers for both processes can be easily overcome by the thermal fluctuations. Hence, at high temperature (i. e. greater than 300 K), more MVs would overcome the energetic barrier to move to the edge region, and finally merge and disappear at the edge. Our results are consistent with the experimental observations showing the formation and growth of void structures at the edge of phosphorene at high temperatures (around 600 K).[28]

**Conclusions**

Phosphorene is well-known for its atomic deficiency due to its relatively low formation energy and highly mobile characteristics. In this work, we performed first-principles and kMC simulations to explore the annihilation and evolution of atomic vacancies in phosphorene. We show that isolated MVs are highly mobile, and prefer to form MV pairs. Under a high MV concentration, the MVs diffuse fast and form MV pairs, which have a much lower mobility and evolve very slowly (in years) under ambient condition. Our calculations also show that while a MV possesses a much lower energy at the edge site, the annihilation process from the interior part to the edge part is quite slow due to a relatively high diffusion barrier. Therefore, MVs are largely retained within the inner part of the sheet under ambient condition. These relatively stable MV clusters may be explored for its application in defective activated catalysis for hydrogen evolution reaction. Our findings may also provide some hints for the repairing of phosphorene structures through chemical functionalization of these vacancy clusters.

**Acknowledgements**

This work was supported in part by a grant from the Science and Engineering Research Council (152-70-00017). The authors gratefully acknowledge the financial support from the Agency for Science, Technology and Research (A*STAR), Singapore and the use of computing resources at the National Supercomputing Centre Singapore (NSCC), Singapore.